\documentclass[prl,aps,twocolumn]{revtex4}
\usepackage{graphicx} 
\usepackage[usenames]{color}
\usepackage{amsmath,amssymb}
\usepackage{gensymb}
\usepackage{natbib}
\usepackage{xcolor}
\def\strutdepth{\dp\strutbox}
\def\nw#1{\strut\vadjust{\kern-\strutdepth\vtop to0pt{\vss\hbox to\hsize
{\hskip\hsize\hskip5pt$\leftarrow$\hss\strut}}}{\em #1}}
\usepackage{rotating}

\usepackage{color}
\definecolor{red}{rgb}{1,0,0}

\usepackage{blindtext}

\begin{document}

\title{Directional shear-jamming}

\author{Martin Trulsson}
\affiliation{Theoretical Chemistry, Lund University, Sweden}

\begin{abstract}
In this work we, study shear reversals of dense non-Brownian suspensions composed of cohesionless elliptical particles. By numerical simulations, we show that a new fragility appears for frictionless ellipses in the flowing states, where
 particles can flow indefinitely in one direction at applied shear stresses but shear-jams in the other direction upon shear stress reversal. This new fragility, absent in the isotropic particle case, is linked to the directional order 
 of the elongated particles at steady shear and its reorientation at shear stress reversal, which forces the suspensions to pass through a more disordered state with an increased number of contacts in which it might get arrested. 
 \end{abstract}
\pacs{83.80.Hj,47.57.Gc,47.57.Qk,82.70.Kj}

\date{\today}
\maketitle

Both granular matter and dense suspensions have been studied extensively during the last decades due to their industrial and geological relevance and rich physics \cite{Andreotti13}. Especially the rheology of these particle packings under various circumstances is currently a hot topic with many questions unresolved, including the exact divergence of the viscosity close to shear-jamming and the associated universality class \cite{Lerner12, Olsson19}, and how to formulate a statistical toolbox for zero-temperature and amorphous states \cite{Baule18}, in analogue with the statistical thermodynamic tool box for equilibrium systems. Even the simplest case of a suspension composed of only repulsive particles at zero temperature shows a vibrant flora of rheological phenomena \cite{Denn14,Guazzelli18}, \emph{e.g.}~transient shear banding \cite{Fielding14} and discontinuous shear-thickening \cite{Seto13,Wyart14,Dong17,Dong20a}. \\
One remarkable property of suspensions is that even though the particles flow in a Stokes flow, particles do not always respect reversibility when exposed to shear reversal \cite{Pine05,Corte08}, a property usually associated with Stokes flows \cite{Taylor00}. 
These irreversible particle trajectories, seen in suspensions after strains typically of the order of unity after a shear reversal, lead to a diffusive motion of particles in oscillatory shear flows with large strain amplitudes. For smaller oscillatory strain amplitudes, particles land in self-absorbed states \cite{Corte08}, showing perfect reversibility. Naturally, there exists a density-dependent critical strain that delimits the reversible-to-irreversible dynamics \cite{Souzy16, Das20}.
The effect is linked to the collisions between the suspended particles, which distort the particles' trajectories from the fluid streamlines. The self-absorbed states correspond to states where the particles restructure themselves to avoid further collisions and hence follow the fluid's streamlines (\emph{i.e.}~show reversibility). 
While first discovered for semi-dilute suspensions, this mechanism has recently been re-discovered for dense suspensions \cite{Lin15, Ness18, Dong20b}, leading to de-thickened states at low oscillatory shear strains. These de-thickened states are naturally linked to the self-absorbing states, as both correspond to zero or very few particle collisions. 
Small oscillatory strains seem to deactivate frictional forces, leading to higher shear jamming packing fractions for frictional particles \cite{Dong20b}. The deactivation of frictional forces is perhaps more evident when one performs shear reversal experiments 
on dense suspensions \cite{Lin15}, where it takes a finite strain, typically around one, before the suspension has restructured and activated the frictional contacts again.\\
Shear-jammed suspensions also show a finite strain over which the suspensions are flowable in the reverse direction \cite{Seto19}, illustrating the fragility of these shear-jammed configurations. This critical strain varies with the packing fraction, starting at around unity close to the shear-jamming packing fraction and decreasing as the packing fraction increases. \newline
Even if fragility has been explored in the jammed region, it has never been seen in the flowing state \emph{i.e.}~below shear-jamming. Here we show that a new kind of fragility appears for suspensions composed of elliptical particles below the steady shear-jamming. This fragility is a new reversible-irreversible transition acting on a suspension as a whole, where a suspension flow { becomes irreversible in the sense that it flows} indefinitely in one shear direction but shear-jams in the opposite direction, \emph{i.e.}~mechanical stable in one direction but not the other, and hence different compared to both the fragility studied in Refs.~\cite{Bi11,Zhao19}, where the authors studied frictional particles starting from stress-free samples and above the corresponding systems yield stress and the undirected reversible-irreversible transition of the dynamics of isotropic particles \cite{Pine05} in oscillating shear-flows. \\
 \begin{figure*}[!htbp]
\centering
\includegraphics[width=0.24\textwidth]{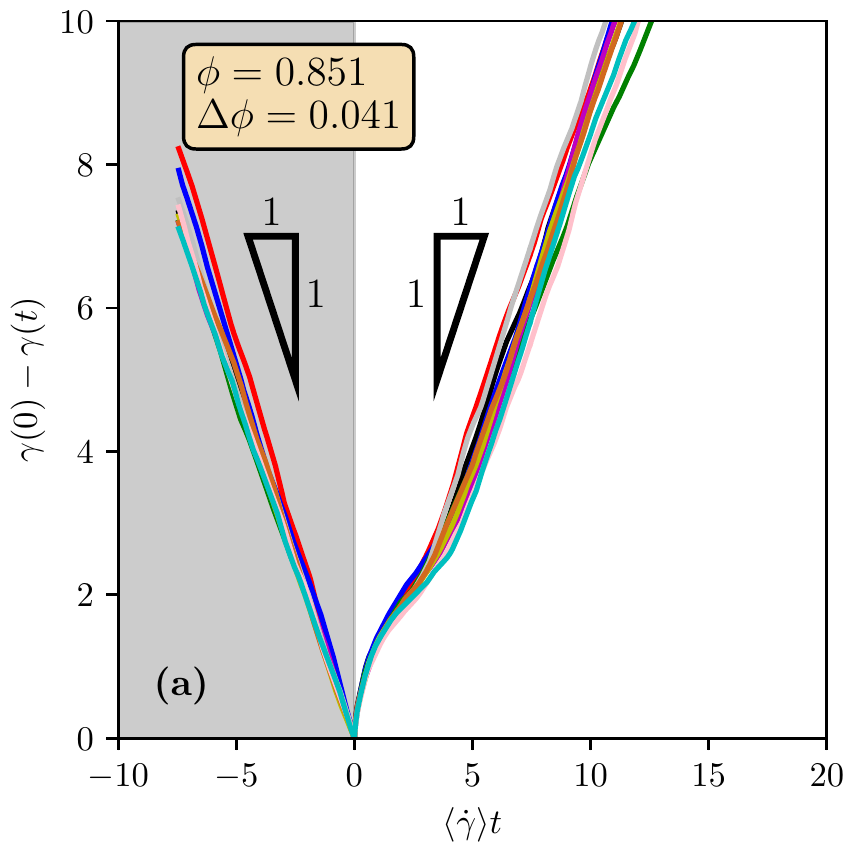}
\includegraphics[width=0.24\textwidth]{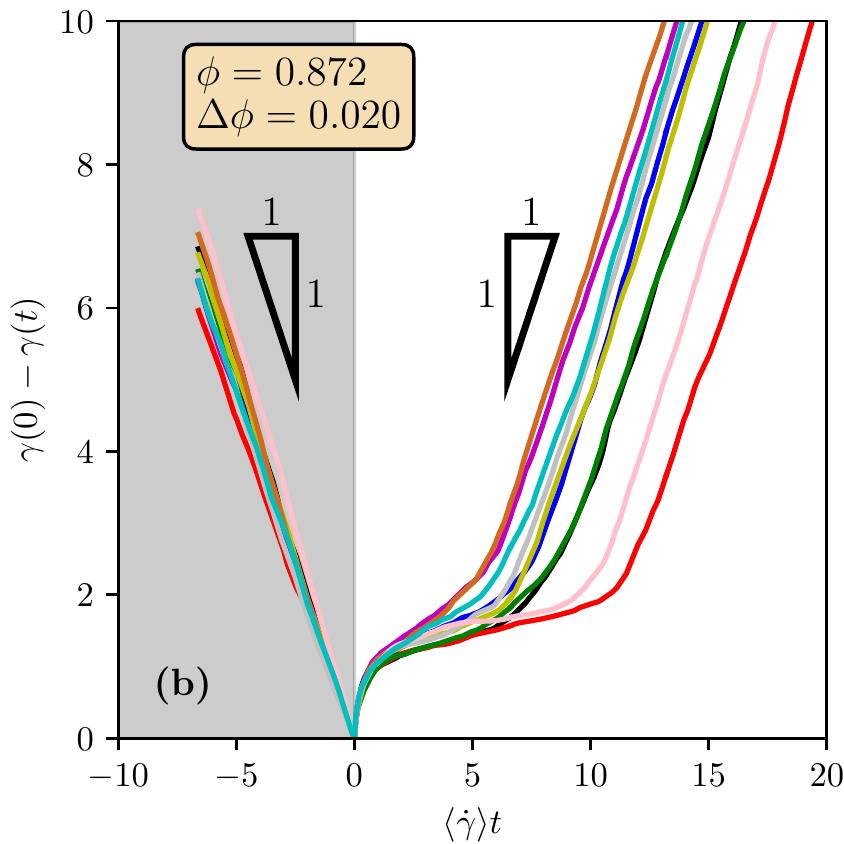}
\includegraphics[width=0.24\textwidth]{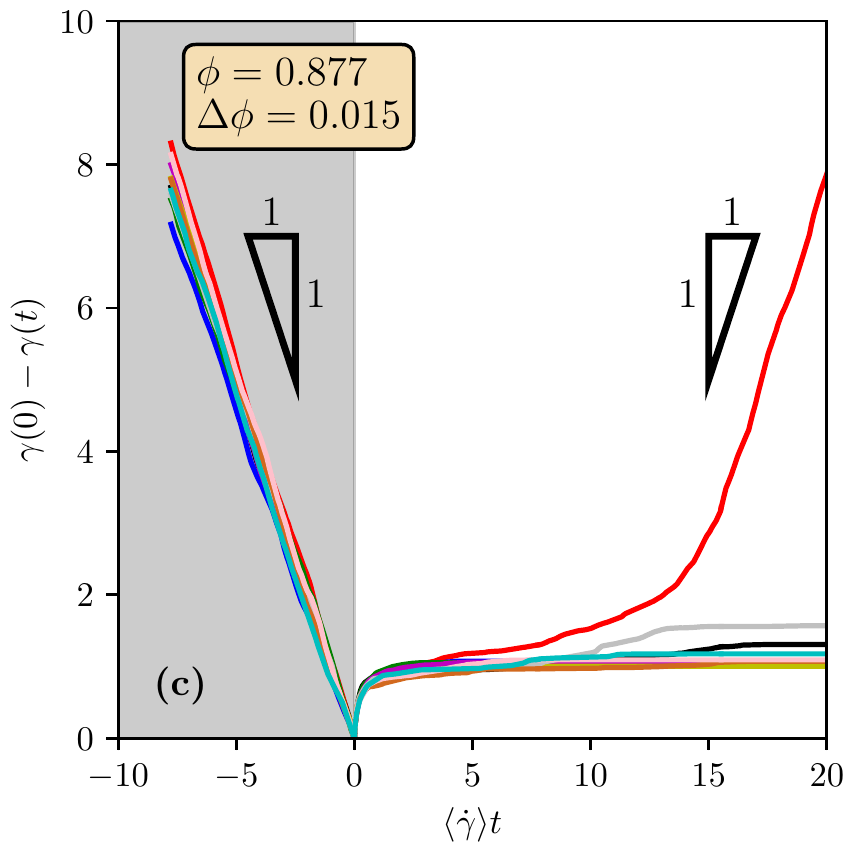}
\includegraphics[width=0.24\textwidth]{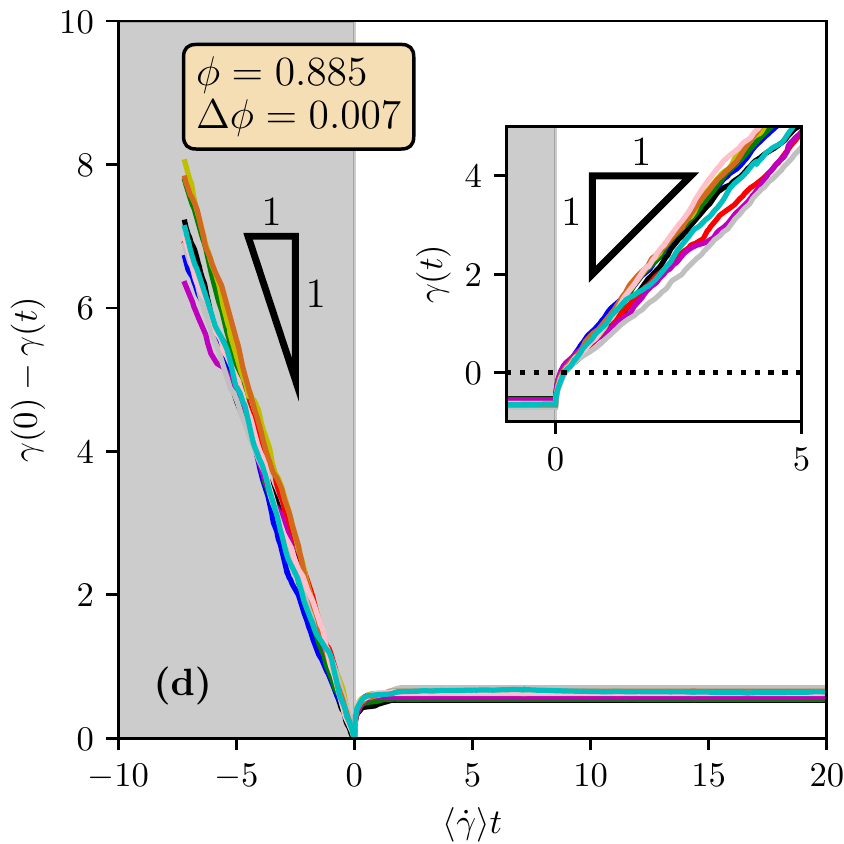}
\includegraphics[width=0.4\textwidth]{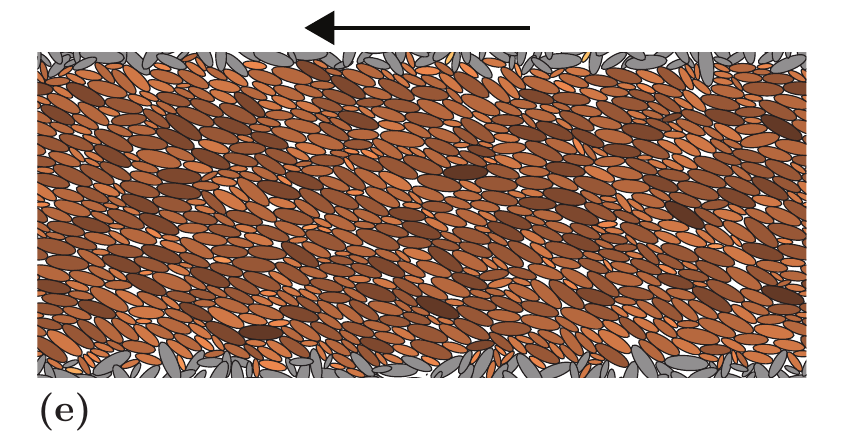}
\includegraphics[width=0.4\textwidth]{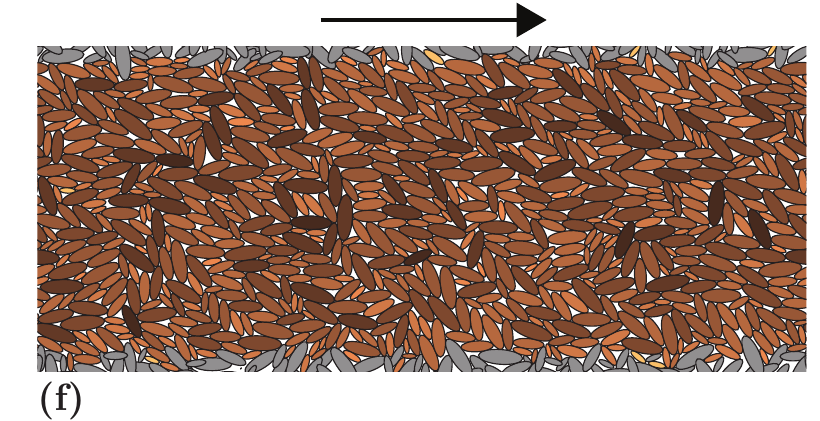}
  \caption{Strain $\gamma$ as a function of the rescaled time $\langle \dot \gamma\rangle t$ for ten independent shear-stress reversals (indicated by various colours). The grey-shaded area indicates pre-reversal evolution and the white post-reversal (upon an instantaneous shear stress reversal, $\sigma \to -\sigma$). Sub-figures show the effect of increasing the packing fraction, from {\bf (a)} to {\bf (d)}, with $\Delta \phi=\phi_c-\phi$. Proportionality triangles show the slopes -1 (left) and 1 (right), respectively.  {\bf (e)} A typical configuration for a flowing state and {\bf (f)} that of a shear reversed jammed state at $\phi=0.877$. Arrows indicated the direction of the shear. Grey particles are wall particles and brown ones flowing particles { (the darker the more contacts)}. Inset of {\bf (d)}  shows the strain evolution upon a second shear reversal from the jammed configurations in the main figure.}
  \label{fgr:AccS}
\end{figure*}
\newline
We carry out extensive numerical simulations of non-Brownian suspensions composed of frictionless elliptical particles (in 2D) in planar shear flows. The particles interact via simple harmonic potentials, where the force is proportional to the overlap. The particles are further subjected to viscous drag forces and torques (see the SI \cite{SI}), the latter giving rise to Jeffrey orbits where the particle angular velocity depends on the orientation. 
We study ellipses with various aspect ratios $\alpha$, defined as the ratio between the lengths of the major $a$ and minor $b$ axes, from 1 (discs) to 3 and at packing fractions up to their respectively shear-jammed states. A typical simulation consists of $\sim900$ particles which all have the same aspect ratio $\alpha$ but various sizes. We shear the particles between two rough walls constructed from the same kind of particles as the flowing particles, randomly oriented. These walls are rough enough to avoid particle slip.  We shear the suspensions at fixed packing fractions by applying shear stress on the two walls with a stress difference denoted by $\sigma$, which are sheared for roughly ten strains{, sufficiently long enough to reach steady-state,} before we do a shear stress reversal (by instantaneously flipping the sign of $\sigma$). In total, we sample ten independent shear-reversals per packing fraction $\phi$. More details about the model{, including the effect on adding lubrication forces or changing the wall roughness,} are provided in the Supplementary Information \cite{SI}.

 \begin{figure*}[!htbp]
\centering
\includegraphics[width=0.24\textwidth]{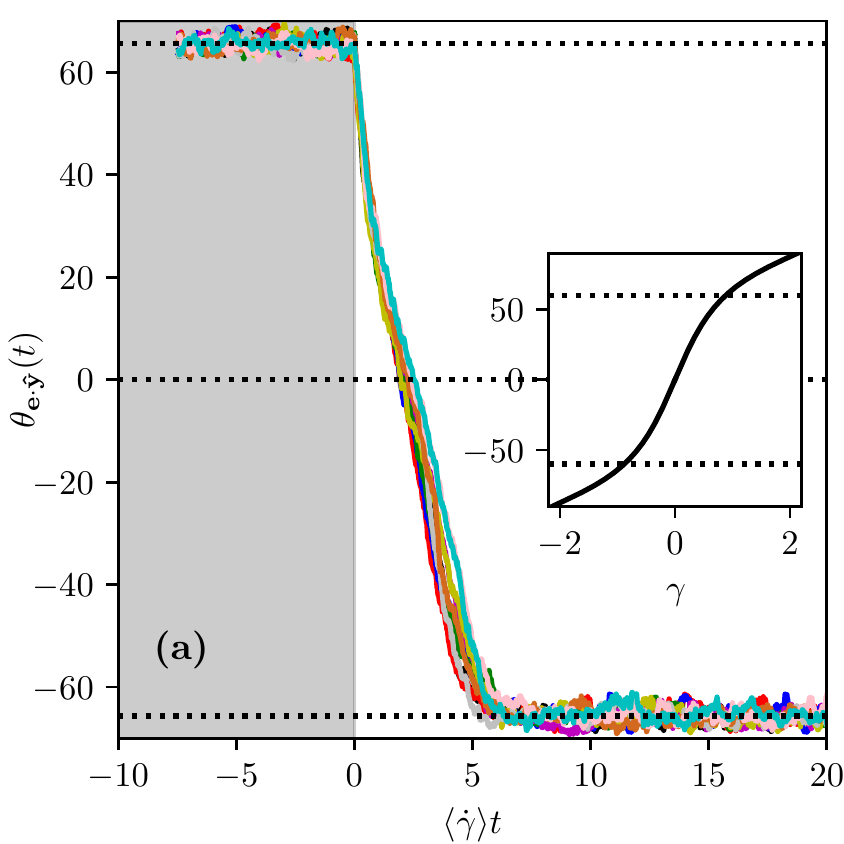}
\includegraphics[width=0.24\textwidth]{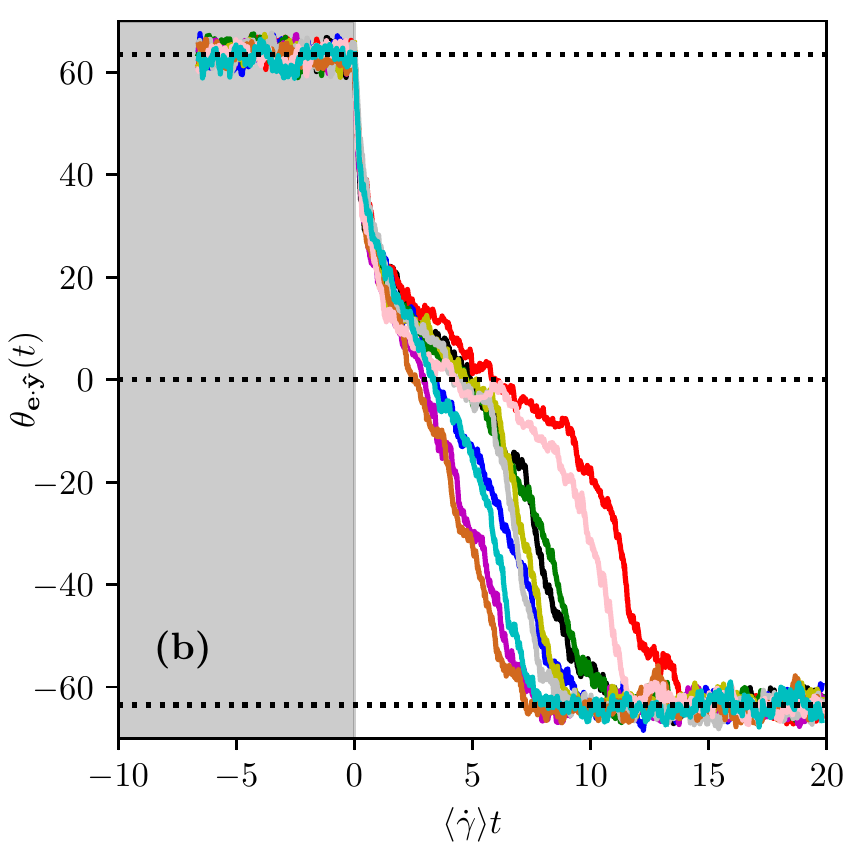}
\includegraphics[width=0.24\textwidth]{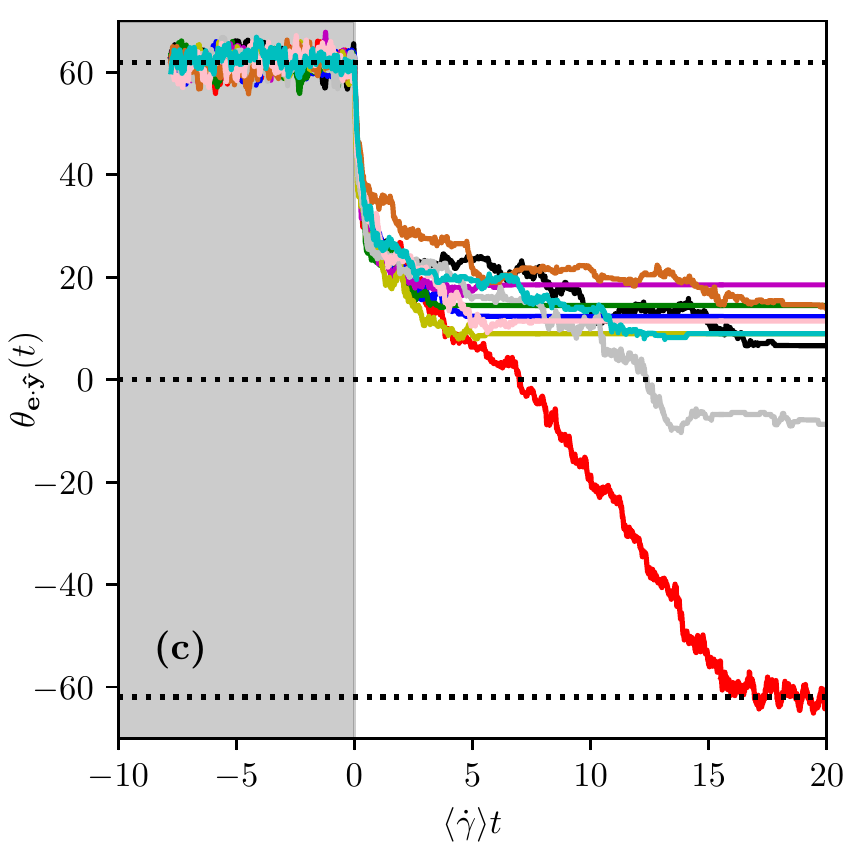}
\includegraphics[width=0.24\textwidth]{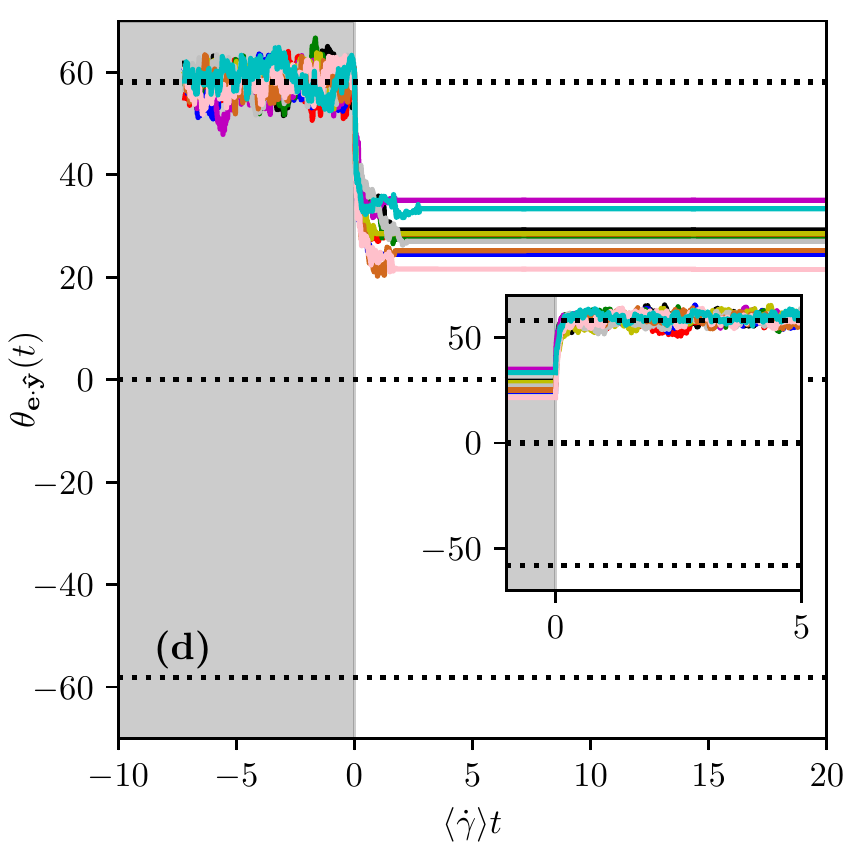}

  \caption{Average particle direction $\theta_{\bf e \cdot \hat{y}}$ (in degrees $^o$) with respect to $\bf \hat{y}$ before (grey-shaded area) and after (white area) shear reversal. Packing fractions {\bf (a-d)} as in Fig.~\ref{fgr:AccS}. Lines show the average
  direction of the ellipses $\pm \langle \theta_{\bf e \cdot \hat{y}} \rangle$ in steady shear and zero. Inset in {\bf (a)} shows the evolution of the angle of a single ellipse in a shear flow as a function of strain. Doing a turn from $-60^o$ to -$60^o$ (indicated by the dashed lines), passing through $0$, requires roughly a strain of 2. Inset in {\bf (d)} as for Fig.~\ref{fgr:AccS} {\bf (d)} but for the average particle direction evolution. 
  }
  \label{fgr:dir}
\end{figure*}
 \begin{figure*}[!htbp]
\centering
\includegraphics[width=0.24\textwidth]{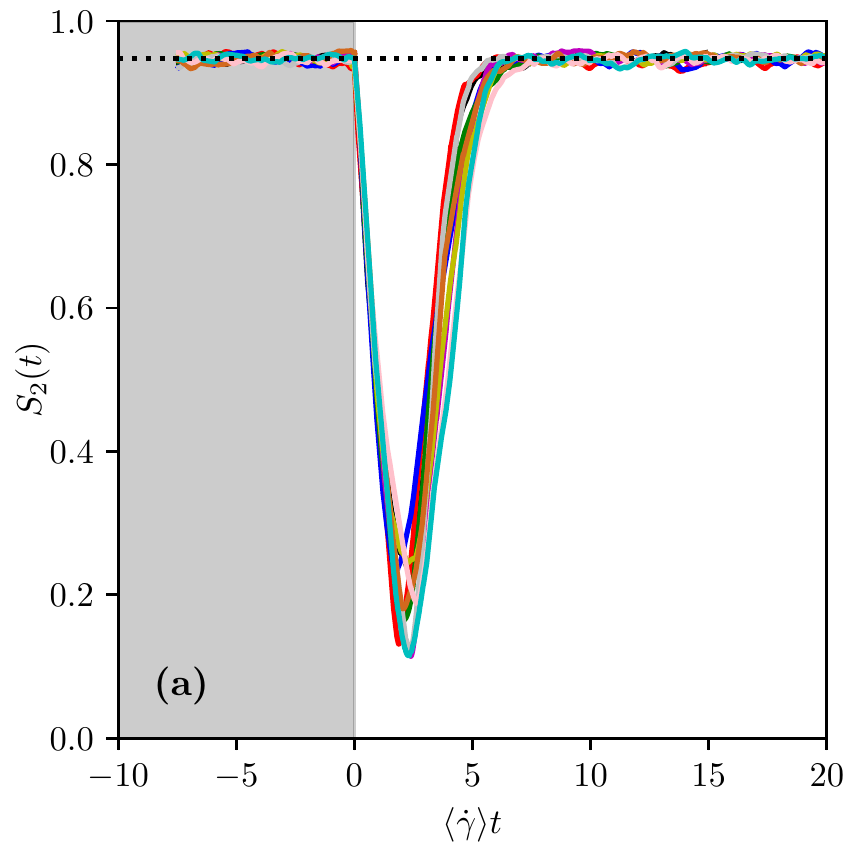}
\includegraphics[width=0.24\textwidth]{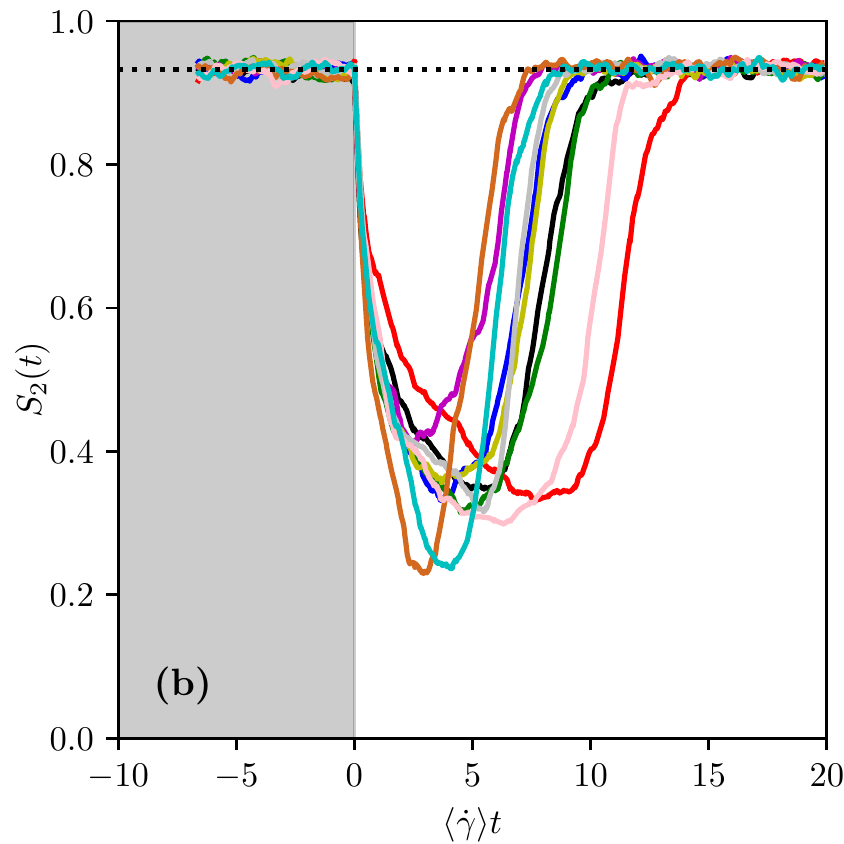}
\includegraphics[width=0.24\textwidth]{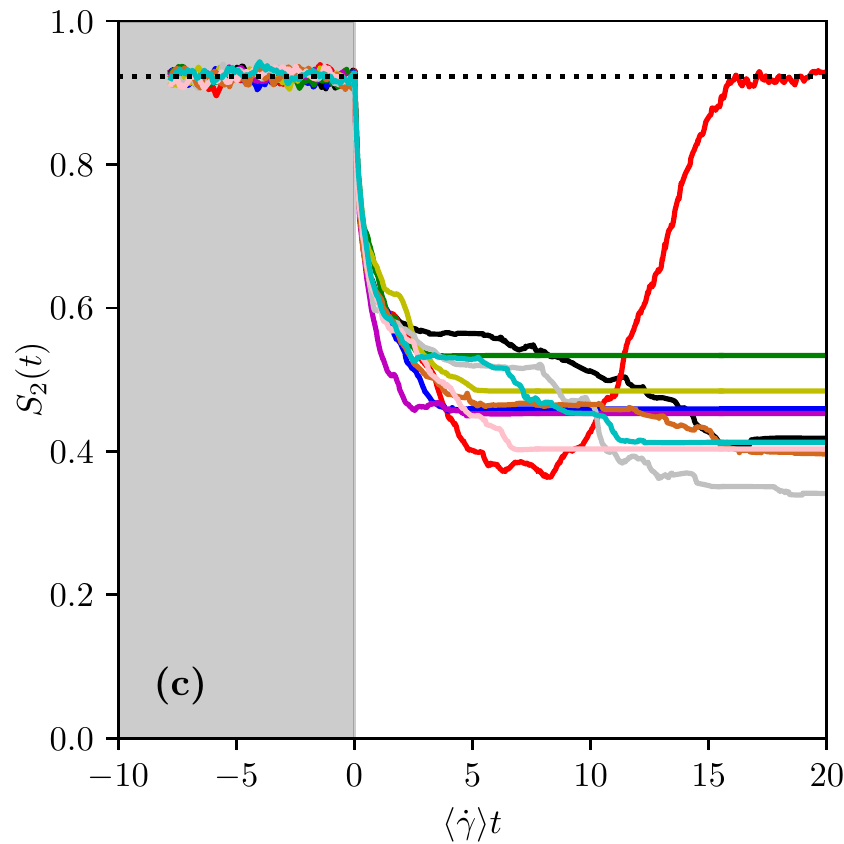}
\includegraphics[width=0.24\textwidth]{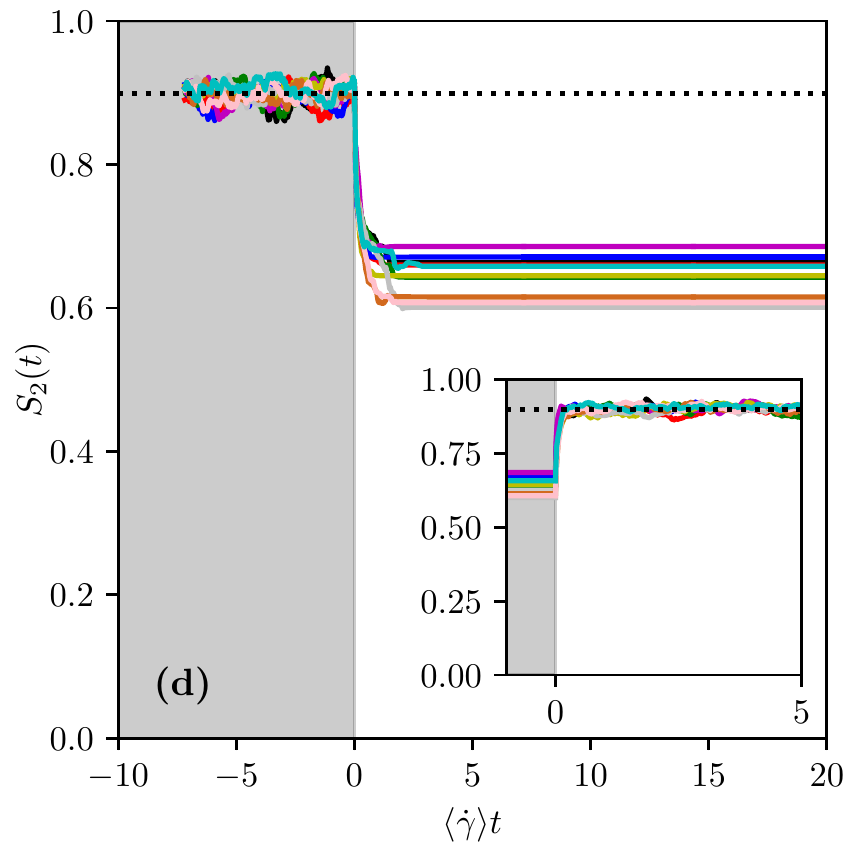}
\caption{Nematic order parameter $S_2$ before (grey-shaded area) and after (white area) shear reversal. Packing fractions {\bf (a-d)} as in Fig.~\ref{fgr:AccS}. Lines show the corresponding averaged value $\langle S_2 \rangle$ as obtained in the steady-state. Inset of {\bf (d)} as for Fig.~\ref{fgr:AccS} {\bf (d)}  but $S_2$.}
  \label{fgr:nem}
\end{figure*}

The key finding is reported in Fig.~\ref{fgr:AccS}, where we report the strain $\gamma$ for ten independent shear stress reversals at four different packing fractions below the shear jamming packing fraction for ellipses, estimated
to be equal to $\phi_{c}=0.892$ for $\alpha=3$ \cite{Trulsson18}. For packing fractions considerably below $\phi_c$, see Fig.~\ref{fgr:AccS}(a) and (b), both the pre-reversal and the post-reversal regions show strains that scale linearly in time with its average steady shear-rate $\langle \dot \gamma \rangle$, as expected for steady flows. After the shear stress reversals, the strains show deviations from a linear scaling, indicating some relaxation process involved. Furthermore, this region consists of two regions: one with non-linear behaviour occurring up to strains of $\sim1$ followed by a quasi-linear regime up to a strain of $\sim 2$. In general, the shear-rate changes sign almost immediately after shear stress reversal. Increasing the packing fraction further, shear-jammed states start to appear upon shear reversal, see Fig.~\ref{fgr:AccS}(c) and (d).
For $\phi=0.877$, nine out of ten realisations get arrested in shear-jammed states upon shear reversal after a strain of roughly one. For an even higher packing fraction, none of the ten systems flows in the reverse direction. 

Even if the packings get arrested in the reverse direction, they immediately start to flow if the stress switches sign a second time, see the inset of Fig.~\ref{fgr:AccS}(d).
These flows' strain evolutions, from slow to arrested at increasing density, after a shear reversal from a flowing state, resembles the typical mean-squared displacements curves and their density dependence for hard-sphere glassy systems \cite{Berthier09}. 
The differences compared to a glassy system are that 1) this effect occurs in an athermal system and 2) this is a macroscopic/collective rheological measurement rather than averaged one particle quantity.\\
Nevertheless, there exist apparent similarities with glassy particle systems, both being arrested systems. Two typical configurations, one of a flowing state and one of an arrested state (upon shear reversal), at $\Delta \phi=\phi_c-\phi=0.015$, are depicted in Fig.~\ref{fgr:AccS}(e-f). By visual inspection, we see a higher directional order characterising the flowing state than the arrested one. 
\\
 \begin{figure*}[!htbp]
\centering
\includegraphics[width=0.24\textwidth]{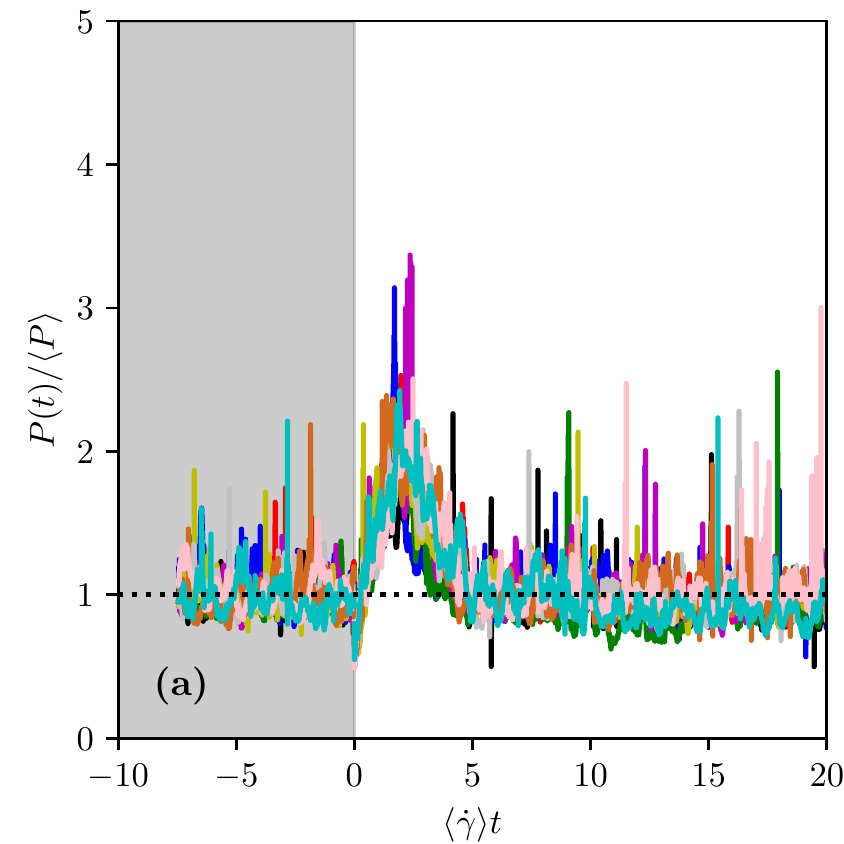}
\includegraphics[width=0.24\textwidth]{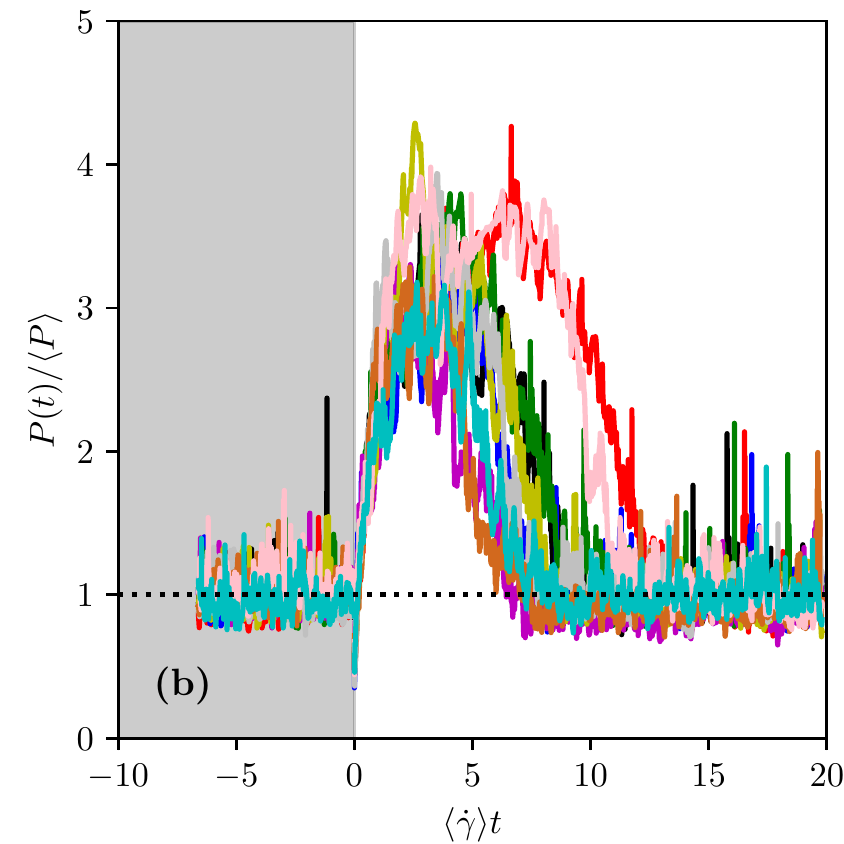}
\includegraphics[width=0.24\textwidth]{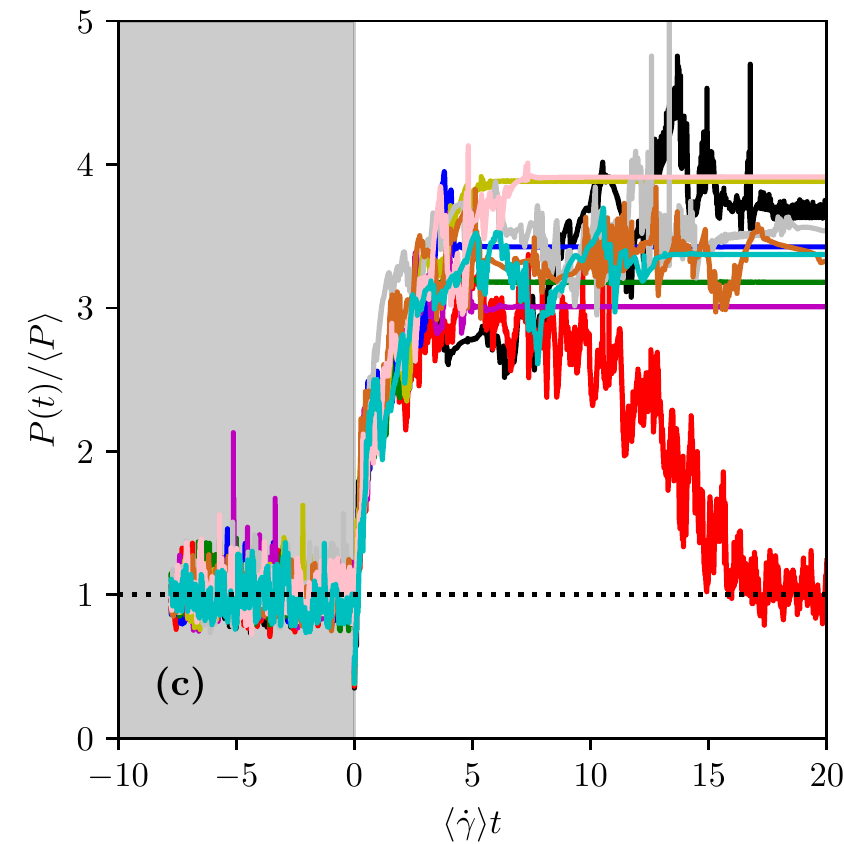}
\includegraphics[width=0.24\textwidth]{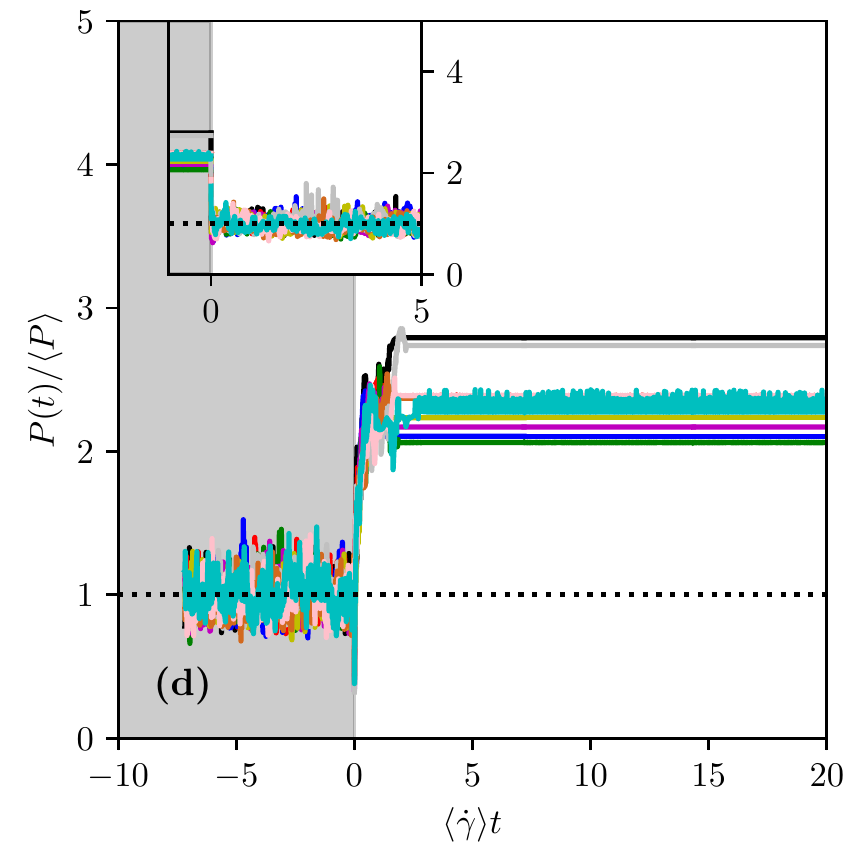}
  \caption{Instantaneous rescaled pressure $P/\langle P \rangle$ at the two walls as a function of the rescaled time $\langle \dot \gamma\rangle t$. Packing fractions {\bf (a-d)} as in Fig.~\ref{fgr:AccS}. Dashed lines show the average pressure $\langle P \rangle$ in steady-state. Inset of {\bf (d)} as for Fig.~\ref{fgr:AccS} {\bf (d)}, \emph{i.e.~}for a second shear reversal, but $P/\langle P \rangle$ .}
  \label{fgr:press}
\end{figure*}

 \begin{figure*}[!htbp]
\centering
\includegraphics[width=0.24\textwidth]{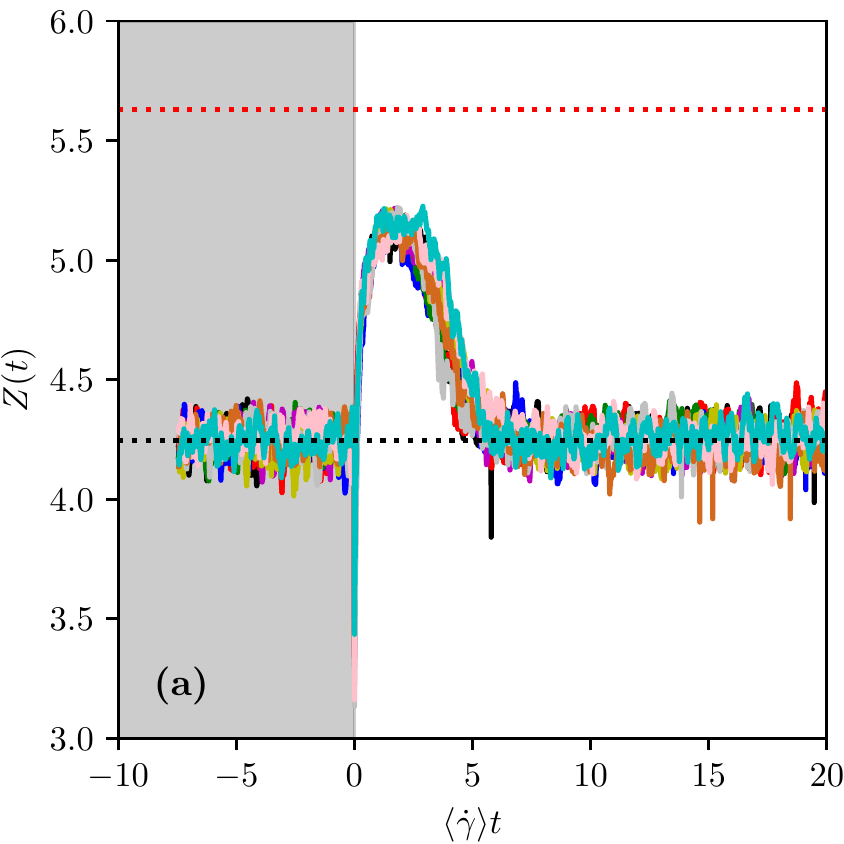}
\includegraphics[width=0.24\textwidth]{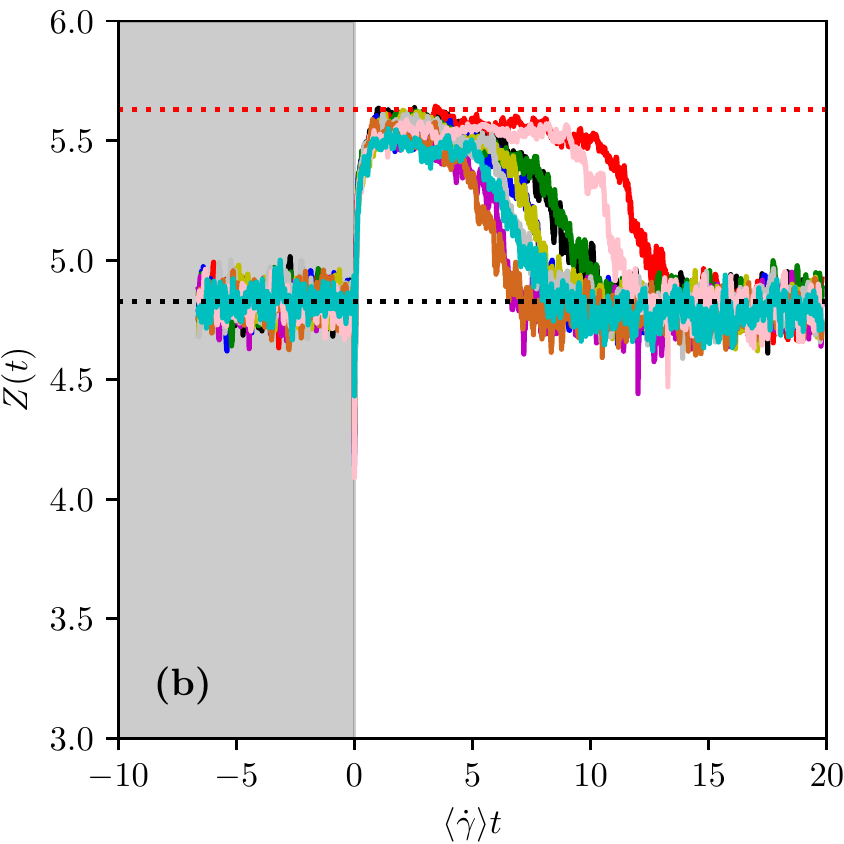}
\includegraphics[width=0.24\textwidth]{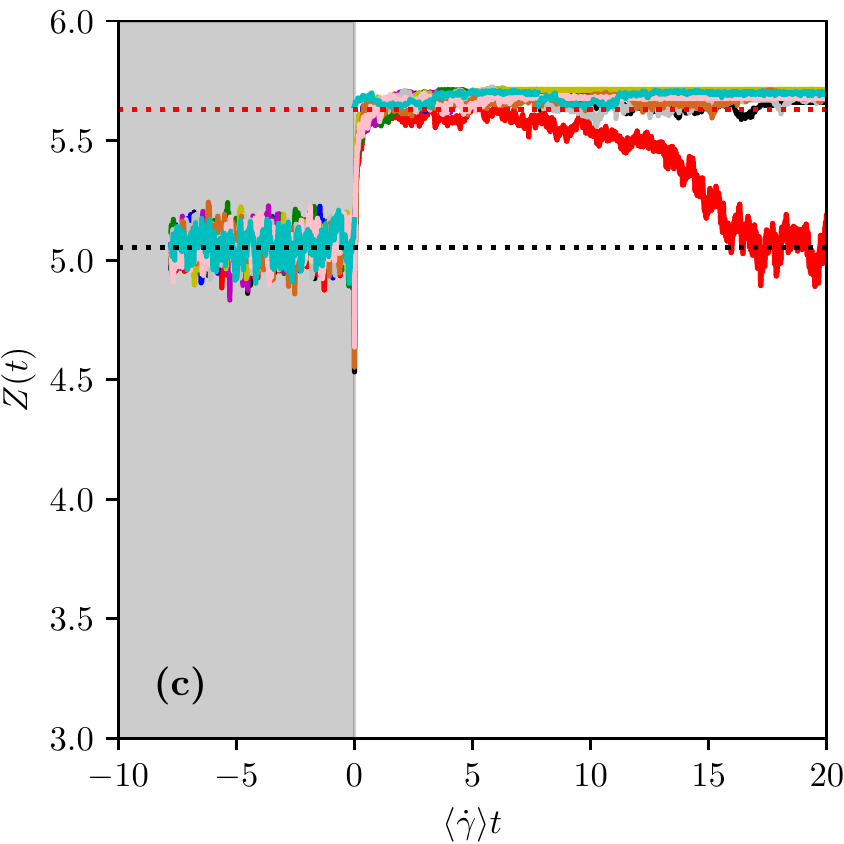}
\includegraphics[width=0.24\textwidth]{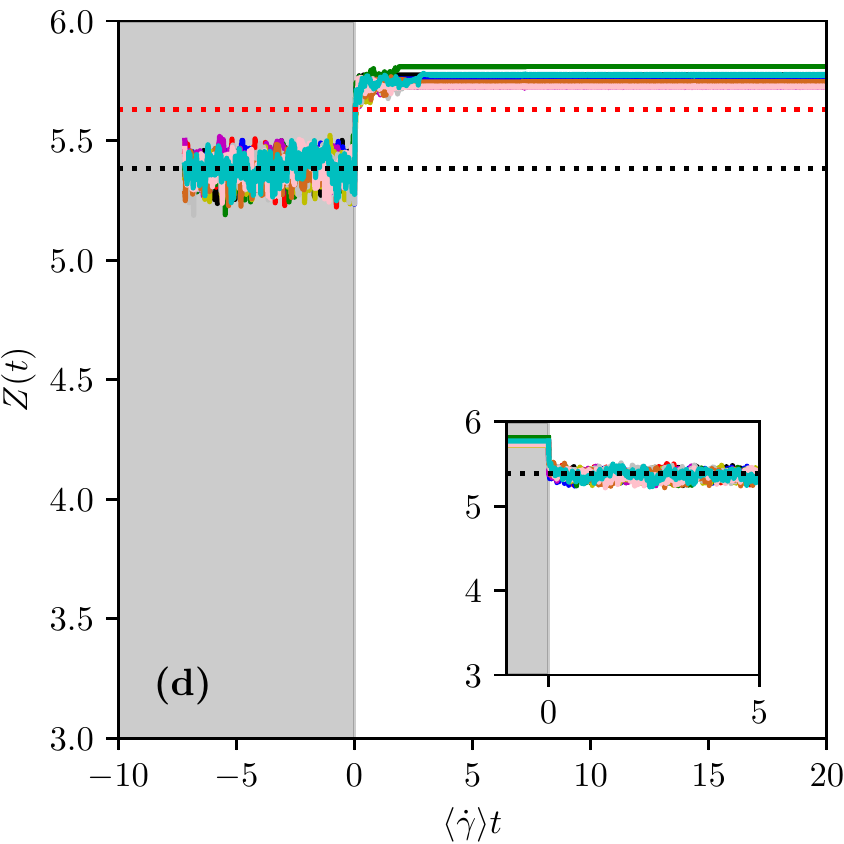}
  \caption{ Instantaneous number of contacts per particle $Z$ as a function of the rescaled time $\langle \dot \gamma\rangle t$. Packing fractions {\bf (a-d)} as in Fig.~\ref{fgr:AccS}. Dashed black lines show the average values of $\langle  Z \rangle$ in steady-state. Dashed red lines show the critical value at steady-state shear-jamming (value from Ref.~\cite{Trulsson18}). Inset of {\bf (d)} as for Fig.~\ref{fgr:AccS} {\bf (d)}, \emph{i.e.~}for a second shear reversal, but $Z$.}
  \label{fgr:contact}
\end{figure*}
\begin{figure}[!htbp]
\centering
\includegraphics[width=0.5\textwidth]{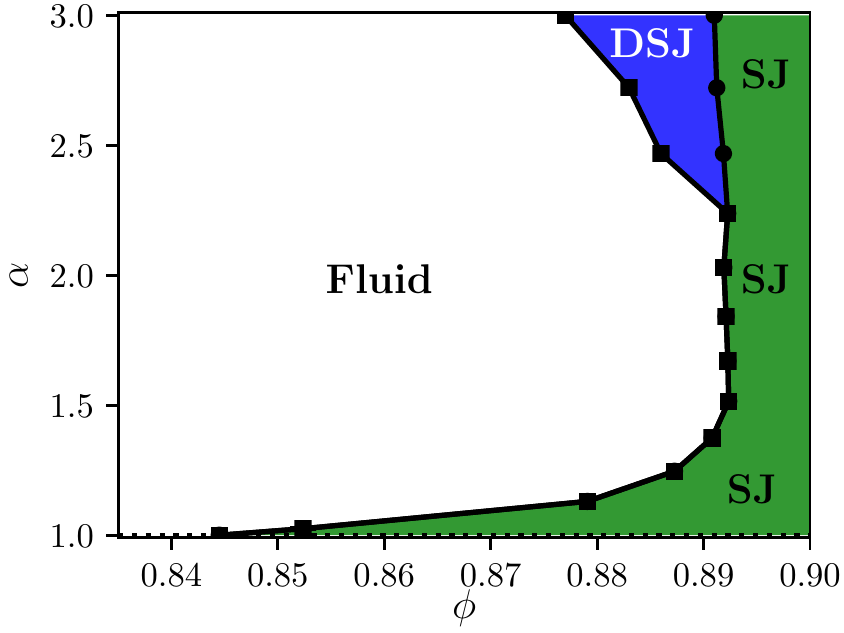}
\caption{Jamming phase diagram in the $\alpha$-$\phi$ plane. The diagram shows the fluid, directional shear jamming (DSJ), and shear jamming (SJ) regions.}
\label{fgr:phi_dsj}
\end{figure}

To better grasp the microscopic origin of these newly found arrested states, and motivated by our above visual inspection, we analyse the systems in terms of directional ordering, both average direction and nematic order parameter of the particles. 
Fig.~\ref{fgr:dir} shows the instantaneous angle (in degrees), averaged over the particles in the centre of the cell, before and after the shear stress reversal. In the pre-reversal regime, the particles' average orientation with respect to the $y$-axis $\theta_{\bf e \cdot \hat{y}}$ fluctuates around its average steady-state value $\langle \theta_{\bf e \cdot \hat{y}} \rangle$, close to $60^o$ for this particular system. At the lowest reported $\phi$, see Fig.~\ref{fgr:dir}(a), the direction slowly switches sign upon reversal and saturates to its negated value, $-60^o$, at around $\langle \dot \gamma \rangle t\approx 5$. The saturation in Fig.~\ref{fgr:dir}(a) correlates well to the re-appearance of linear scaling of the strain with respect to time, see Fig.~\ref{fgr:AccS}(a). Hence, the deviation from linear scaling in the post-reversal region can be attributed to the slow reorientation of the direction of the ellipses to the new reversed stress direction. In terms of strain, this completed reorientation corresponds to approximately 2 in the reverse direction, extracted from the data in Fig.~\ref{fgr:AccS}(a). A single ellipse with $\alpha=3$ in a shear flow has a turn-over strain roughly equal to 4.3. However, a $60^o$ to $-60^o$ reorientation corresponds roughly to 2 strains, see the inset of Fig.~\ref{fgr:dir}(a), and comparable to what we find in the simulations.
This finding is similar to what found in 3D systems of anisotropic particles at lower packing fractions using a constant shear-rate rheometer \cite{Nadler18}, even though the relaxation is much faster in those cases, with reported values around 0.2 in strain. This smaller value merely reflects a smaller tilt angle in those cases (\emph{e.g.,}~$\pm 20^o$). 
We see similar trends increasing the packing fraction, where the linear scaling and a fully reoriented direction of the ellipses coincide. This observation is valid for all the cases where we find a flowing state in the reverse direction.
For the arrested states, most configurations get stuck with an average angle of the same sign as for the pre-reversal shear direction. \\
Further insights can be obtained if one also looks at the instantaneous nematic order $S_2$, as shown in Fig.~\ref{fgr:nem}. Following previous results \cite{Nadler18}, the nematic ordering starts to decrease after a reversal until it reaches a minimum, with a corresponding value of approximately zero for $\theta_{\bf e \cdot \hat{y}}$, after which it increases back to its original value. For the studied cases, we see a nematic order parameter $S_2$ of above 0.9 (high nematic order) at steady shear but reaches as low as 0.2 (low nematic order) after a shear stress reversal. Hence, to reorient the flow in the reverse direction, one passes through a more disordered state (in terms of nematic ordering). While for $\phi=0.877$, the system gets jammed in this disordered state (see Fig.~\ref{fgr:AccS}(f)), with low nematic ordering and no strong preferential direction along with the flow, this is not the case at even higher packing fractions where the system instead gets stuck in a more ordered state { (compared to the $\phi=0.877$ case)} with both a moderate-high nematic order{, but still less ordered than in steady-state,} and a clear anti-alignment compared to the new stress direction.
Hence, these arrested states can reach different degrees of nematic disorder and average particle orientations. \\
We now turn our attention to the pressure evolution upon shear reversal. Fig.~\ref{fgr:press} shows the rescaled instantaneous normal stress on the walls as a function of time at various packing fractions. For the lowest reported packing fraction, we see a small but still significant maximum. This maximum coincides nicely with when the suspension has its minimum in nematic order and $\theta_{\bf e \cdot \hat{y}}\approx 0$, \emph{i.e.} a fairly disordered configuration with no preferential direction in either flow direction. The pressure maximum increases up to roughly 3.5 times the steady-state value at even higher packing fractions. For the highest packing fraction, the pressure saturates to a value lower than this maximum, most likely because
the pressure is not fully developed as the suspension gets arrested before { the most} disordered state is reached. \\
This pressure increase can be understood if one considers the average number of contacts per contact $Z$ and its evolution upon shear-reversal: see Fig.~\ref{fgr:contact}, where the pressure increase well correlates with an increased number of contacts compared to its steady-state value. As the packing fraction increases. so do the average number of contacts per particle, both for steady-state and upon shear-reversal. For steady-state, a suspension with $\alpha=3$ shear-jams when $Z=5.63\pm 0.02$ \cite{Trulsson18}, indicated by red dashed lines in Fig.~\ref{fgr:contact}. Directional shear-jammed states appear when $Z$ reaches or exceeds this steady-state number. Hence, upon shear-reversal, the elliptical particles pass through a more disordered state with a higher number of contacts, possibly leading to shear-jamming. It, furthermore, indicates that the shear-jamming is controlled by the number of contacts per particle rather than the packing fraction and that this value is roughly independent of the value of the nematic order parameter. \\
After seeing the above results, one natural question arises: ``At which aspect ratio does these arrested states appear?''
In order to answer this question, we carry out additional simulations for suspensions composed of particles with lower aspect ratios. 
It turns out that directional arrested states start to appear as soon as the aspect ratio is greater than roughly 2.2 (see Fig.~\ref{fgr:phi_dsj} and the Supplementary Information \cite{SI}).
Interestingly, this is also the reported aspect ratio for which mono-layered colloidal ellipsoids (2D systems) starts to have two distinct glass transitions densities:~a lower for the rotational motion and a higher one for the translational one \cite{Zheng14}.\\
\newline
In summary, we have presented a new type of fragility present in dense non-Brownian suspensions composed of anisotropic particles. We have illustrated this for elliptical particles with an aspect ratio $\alpha=3$.
Even though the suspension behaves as a Newtonian fluid when shear stress is applied in one direction, it shear-jams when the stress is applied in the reverse direction for packing fractions greater than $\phi>(\phi_c^{\alpha=3}-0.015)$.
For ellipses this fragility, or directional shear-jamming, starts to be important for $\alpha>2.2$ and packing fractions close to the steady-state shear-jamming. 
These novel fragile states are fundamentally different from the fragile states seen for isotropic and frictional particles \cite{Bi11} as those systems shear-jam in any direction given that the strain is large enough.
Furthermore, it is not unlikely that spherocylinders \cite{Nagy17} are more prone to show this fragility due to their
more localised curvature as compared to ellipses. Since granular materials and dense non-Brownian suspensions share much physics \cite{Boyer11,Trulsson12,Amarsid17}, it would not come as a surprise if this fragility also exists for granular materials, possibly shifted due to inertial effects. Investigation of this effect for granular material as well as for frictional particles \cite{Trulsson18} will thus be a natural extension of this work. \\
\newline
The author thanks J. Stenhammar for valuable comments. The simulations were performed on resources provided by the Swedish National Infrastructure for Computing (SNIC) at the centre for scientific and technical computing at Lund University (LUNARC).

\section{Supplementary Information}
\subsection{Model and simulations}
We model an amorphous ensemble of 2D quasi-hard ellipses interacting with Hookean, \emph{i.e.}~linear, springs

\begin{equation}
\bold{f}_{ij} = k \boldsymbol{\delta}_{ij}^\perp,
\end{equation}

where $k$ is the spring constant and $\boldsymbol{\delta}_{ij}^{\perp}$ the overlap, normal to the surface, between the two ellipses $i$ and $j$. Since the ellipses are anisotropic and force is normal to the surface, the ellipses are also subjected to torques in relation to the forces. For more information about the contact forces, see \cite{Trulsson18}.
The major axis length of an ellipse is randomly chosen from a flat distribution between $a/\langle a\rangle=[0.5,1.5]$.
Particles are quasi-hard, where the ratio $k/\langle P\rangle$ is set so high (typically above $10^3$) that the obtained results are insensitive to the value of $k$ (within the noise of the data). $\langle P \rangle$ is here the average steady shear pressure (normal stress on the walls).

The fluid is described as a continuum, imposing a linear shear profile with the fluid velocity in the x-direction equal to $u_{f,x}(y)= \dot \gamma y$, where $\dot \gamma$ is the shear rate. The velocity field respects non-slip conditions in the vicinity of the walls, \emph{i.e.,}~$\dot \gamma=\Delta u_{w,x} /H$, where is $\Delta u_{w,x}$ lateral velocity difference and $H$ the separation between the two walls. Besides contact forces, the ellipses experience viscous drag as 
\begin{equation}
\bold{f}_i^{\rm visc}= -3\pi \eta_f \big[ c_{fa}\big(\bold{u}_i^a-\bold{u}_f^a(y_i)\big)+c_{fb}\big(\bold{u}_i^b-\bold{u}_f^b(y_i)\big) \big],
\end{equation}
and torque
\begin{equation}
\tau_i^{\rm visc} = 4 \pi \eta_f a_i b_i \big[ \big(c_{Ma} (e_{i,x})^2+c_{Mb} (e_{i,y})^2\big)\Omega-c_{Mr} \omega_i \big],
\end{equation}
where $e=\sqrt{1-\alpha^{-2}}$ is the eccentricity, $\eta_f$ the interstitial fluid viscosity, $\bold{u}_i^{(a/b)}$ are the translational velocities along the major and minor axes respectively of particle $i$, $\bold{e}_i=(e_{i,x},e_{i,y})$ the unit direction vector of particle $i$'s major axis, $\Omega=\frac{\dot \gamma}{2}$ the vorticity of the fluid, $\omega_i$ the angular velocity of particle $i$, and $c$ are constants:
\[c_{fa} = \frac{8}{3}e^3\big[-2e+(1+e^2)\log\big(\frac{1+e}{1-e}\big)\big]^{-1}, \]
\[ c_{fb} = \frac{16}{3}e^3\big[2e+(3e^2-1)\log\big(\frac{1+e}{1-e}\big)\big]^{-1}, \]
\[c_{Ma}=c_{fa}, \]
\[c_{Mb}=(1-e^2)^{-1}c_{fa}, \]
\[c_{Mr}=\frac{4e^3 (2-e^2)}{3(1-e^2)} \Big[-2e+(1+e^2) \log \big( \frac{1+e}{1-e} \big) \Big]^{-1}. \]
Particle dynamics is overdamped with force and torque balances on each particle. \\
Starting configurations for the constant volume and boundary stress simulations are all prepared by a pressure-controlled rheometer where an external pressure $P_{\rm ext}$ is imposed at the walls and with a constant $\Delta u_{w,x}$, resulting in an almost fixed shear rate after a brief period. The desired packing fraction is then obtained by slowly decreasing or increasing the shear-rate under the imposed pressure until the desired value is reached.
We then perform constant stress simulations at constant packing fractions by locking the wall separation and imposing a constant stress $\sigma$ (equivalent to a constant force) on the top wall and zero on the bottom.
Shear reversal is then done by flipping the sign of $\sigma$. \\

\subsection{Average direction and nematic order parameter}
The average direction with respect to the $y$-axis, with the unit vector $\bf \hat{y}$, is sampled by taking the average orientation of all ellipses in the centre of the cell, excluding the five closest layers close to each wall. The nematic order parameter is obtained from the
instantaneous director tensor $\bold{Q}_{lk}=\frac{1}{N} \sum_{i=1}^N (2 e_{i,l} e_{i,k}-\delta_{lk}) $, where $\delta_{lk}$ is the Kronecker delta and where $l$ and $k$ are either $x$ or $y$. The nematic order parameter was sampled for the $N$ particles in the centre of the cell, as done for the average direction. The nematic order parameter is then obtained as $S_2=\sqrt{ {\bf Q}_{xx}^2+{\bf Q}_{xy}^2}$. \\

\setcounter{figure}{0}
\makeatletter 
\renewcommand{\thefigure}{S\@arabic\c@figure}
\makeatother

\subsection{Shear reversal at lower aspect ratios}
\begin{figure}[!htbp]
\centering
\includegraphics[width=0.22\textwidth]{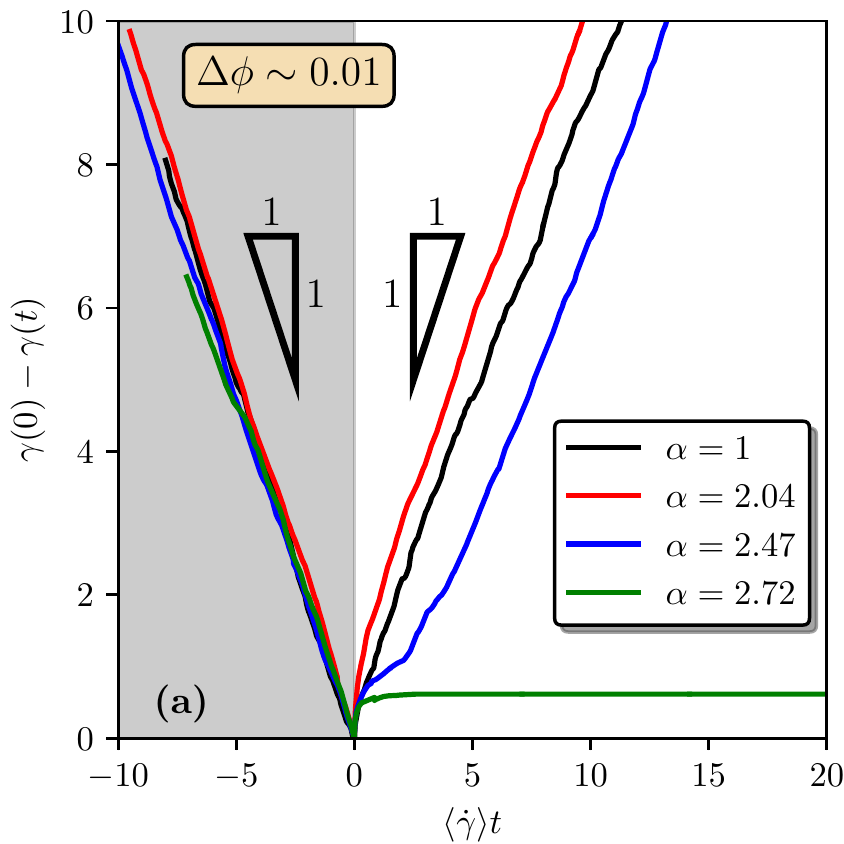}
\includegraphics[width=0.22\textwidth]{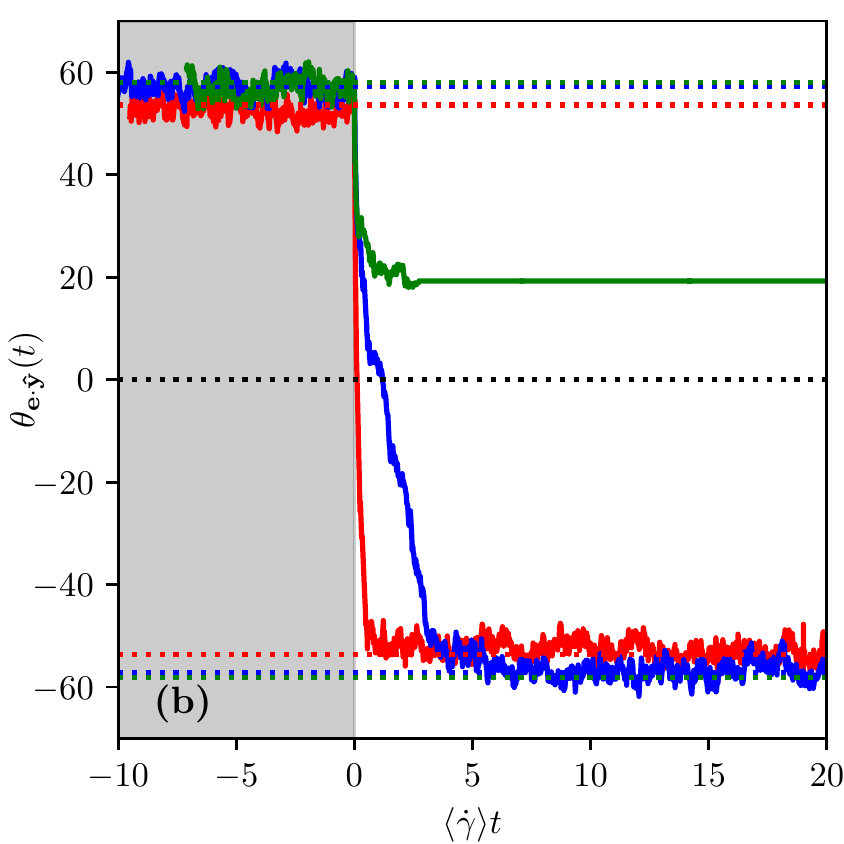}
\includegraphics[width=0.22\textwidth]{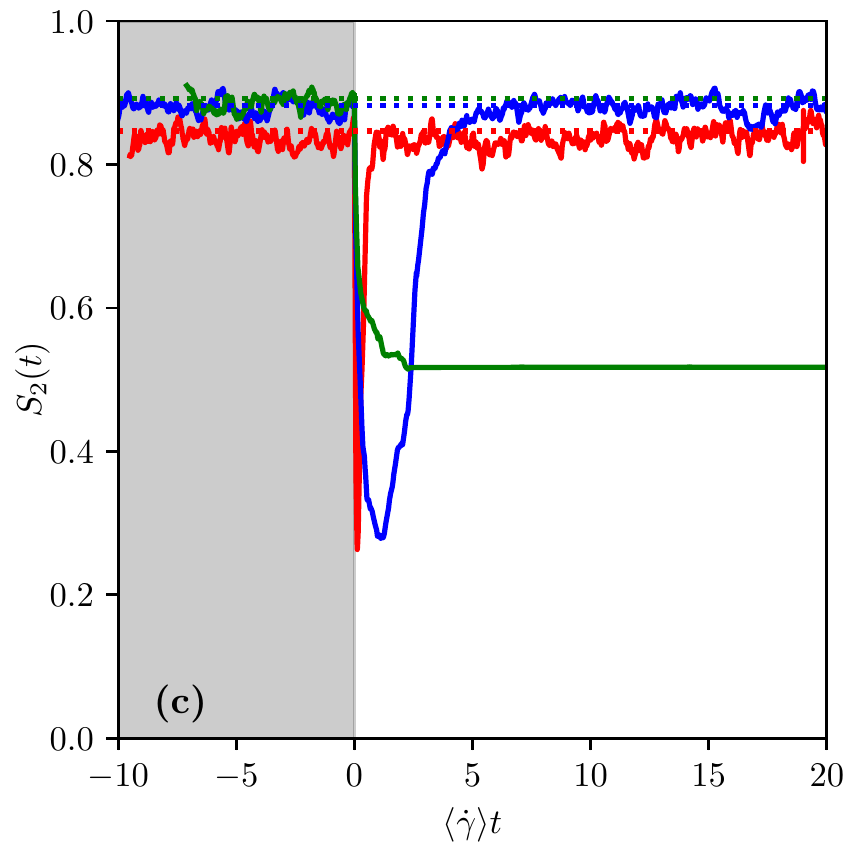}
\includegraphics[width=0.22\textwidth]{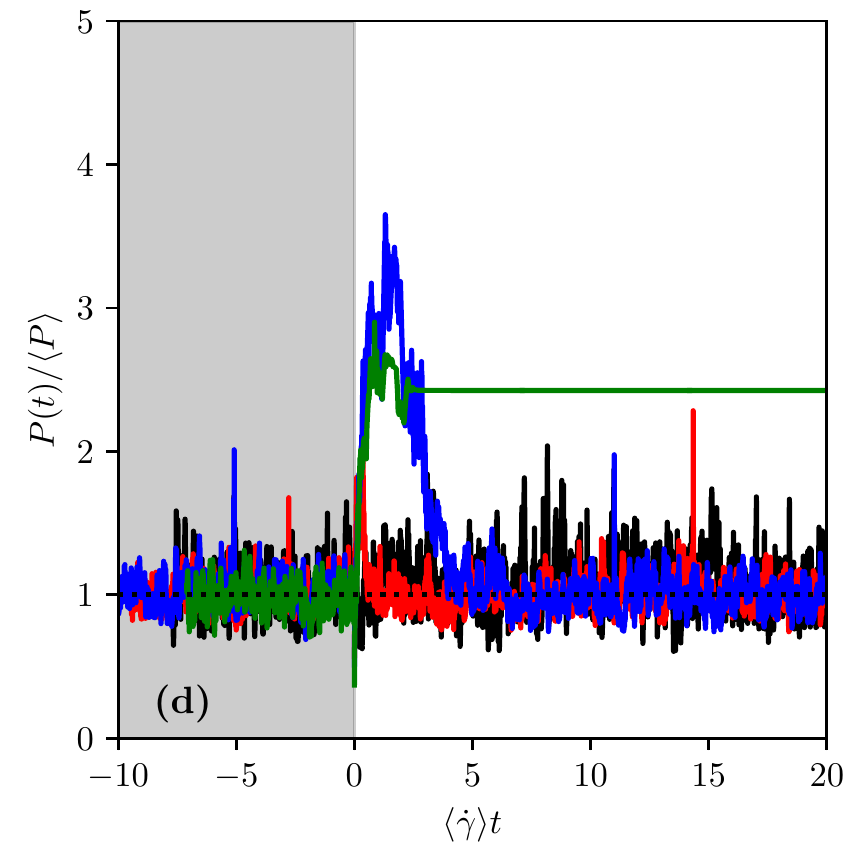}
\includegraphics[width=0.22\textwidth]{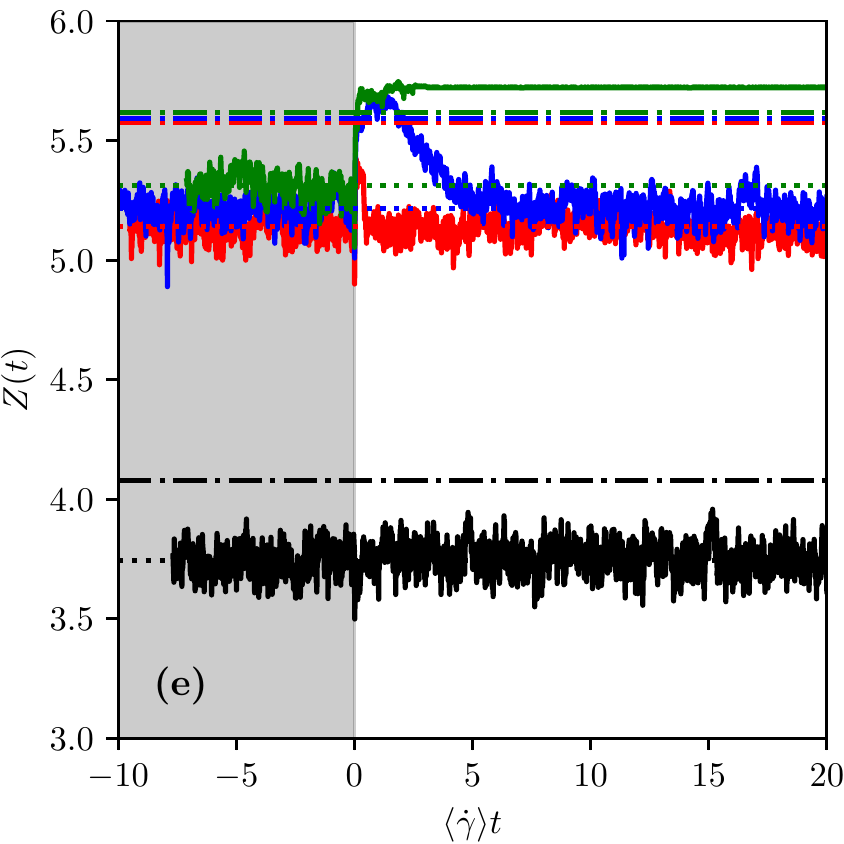}
\caption{{\bf (a)} Strain $\gamma$, {\bf (b)} average particle direction $\theta_{\bf e \cdot \hat{y}}$ (in degrees $^o$) with respect to $\bf \hat{y}$, {\bf (c)} nematic order parameter $S_2$, {\bf (d)} instantaneous pressure at the two walls, rescaled by its corresponding steady shear value, and { {\bf (e)} average number of contacts per particle} as a function of time for four different aspect ratios as indicated by the legend in {\bf (a)}. Dotted lines in {\bf (b)} and {\bf (c)} show the corresponding values at steady shear. }
 \label{fgr:VarA}
\end{figure}

Fig.~\ref{fgr:VarA} shows strain, instantaneous direction angle, nematic order parameter, pressure{, and number of contacts per particle evolutions} for suspensions with aspect ratios $\alpha<3$ at low and comparable relative distances to their respectively shear jamming packing fractions, $\Delta \phi \sim 0.01$. For $\alpha=1$ (disc particles) and $\alpha=2.04$, the linear scaling in the strain as a function of time is almost immediately recovered upon shear reversal without any significant maximum pressure.
The directional angle and nematic order parameter for $\alpha=2.04$ also relax almost immediately. For larger aspect ratios, these relaxations become slower with the appearance of a slow and quasi-linear regime in the strain curves as a consequence, and pressure { and number of contacts} maxima.

{ 
\subsection{Effect of lubrication forces}
To verify that the directional shear-jamming is not diminished by short-range lubrication forces, we carry out additional simulations with lubrication forces included.  The lubrication is modelled as previously done for discs \cite{Dong21}.
The squeeze mode (normal to the surfaces) is given by:

\begin{equation}
\mathbf{f}^{ij}_{\mathrm{lub,n}}= -\frac{3 \pi}{8} \eta_f \kappa_{ij}^{-1} \Big[ \frac{(\mathbf{V}_i-\mathbf{V}_i)\cdot \mathbf{n}_{ij}}{h_{ij}^\perp+\delta_{\rm rough}}\Big] \mathbf{n}_{ij},
\end{equation}

where $\kappa_{ij}^{-1}=\frac{2\kappa_i^{-1}\kappa_j^{-1}}{\kappa_i^{-1}+\kappa_j^{-1}}$ is a reduced inverse curvature using the curvatures, $\kappa_i$ and $\kappa_j$, of the two ellipses at the closest contact points, $\mathbf{n}_{ij}$ the normal unit vector to the two surfaces, $\mathbf{V_i}$ the total velocities (\emph{i.e.,}~including both translational and rotational motions) at the closest contact points of ellipse $i$, $h_{ij}^\perp$ the gap perpendicular to the surface between the two closest points, and $\delta_{\rm rough}$ a roughness parameter. If the gap $h_{ij}^\perp$ is negative (and $\boldsymbol{\delta}^\perp$ non-zero) the lubrication is set equal to zero.  We have equally accounted for the weaker shear mode:

\begin{equation}\label{eq:lubt}
\mathbf{f}^{ij}_{\mathrm{lub,t}} = \Big[-\frac{1}{2}\pi\eta_f\ln\Big{(}\frac{\kappa_{ij}^{-1}}{2(h_{ij}^\perp+\delta_{\rm rough})}\Big{)}(\mathbf{V}_i-\mathbf{V}_j)\cdot\mathbf{t}_{ij}\Big]\mathbf{t}_{ij},
\end{equation}

where $\mathbf{t}_{ij}$ is the tangential vector at the closest contact points. We chose to have the roughness parameter equal to $\delta_{\rm rough}/\sqrt{\langle ab \rangle}=0.025$.\\
Fig.~\ref{fgr:Lubr} shows that including lubrication forces does not diminish the directional shear-jamming. These simulations have been run with Newtonian dynamics but should be considered as overdamped as the 
Stokes number $St = \frac{\rho\dot \gamma \sqrt{ab}}{\eta_f}\ll1$, where $\rho$ is the mass density of particles. 

\begin{figure}[!htbp]
\centering
\includegraphics[width=0.5\textwidth]{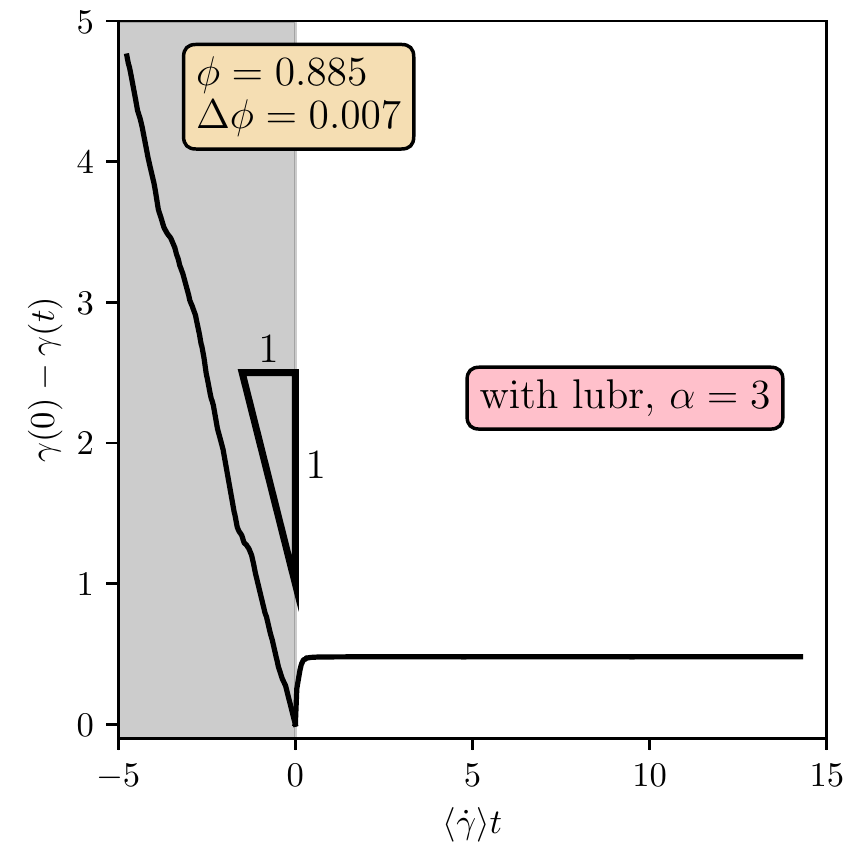}
\caption{ As in Fig.~\ref{fgr:AccS}{\bf (d)} with lubrication forces included ($St=2.5 \cdot 10^{-3}$).}
 \label{fgr:Lubr}
\end{figure}

\subsection{Effect of wall roughness}

In the main manuscript, we altered the wall particles' shapes. In order to explore if the directional shear-jamming is caused by the boundary, we carry out two extra simulations. The first is by altering only the wall particles shape keeping the flowing particles as discs. The other test is doing the reverse, \emph{i.e.,}~keeping the wall particles as discs but changing the flowing particles' shapes. As seen in Fig.~\ref{fgr:wall}, the wall roughness does affect the dynamics and, hence, does not alter the conclusion reached in the main article. Therefore, we can conclude that the directional shear jamming is not a boundary-driven effect. 

\begin{figure}[!htbp]
\centering
\includegraphics[width=0.23\textwidth]{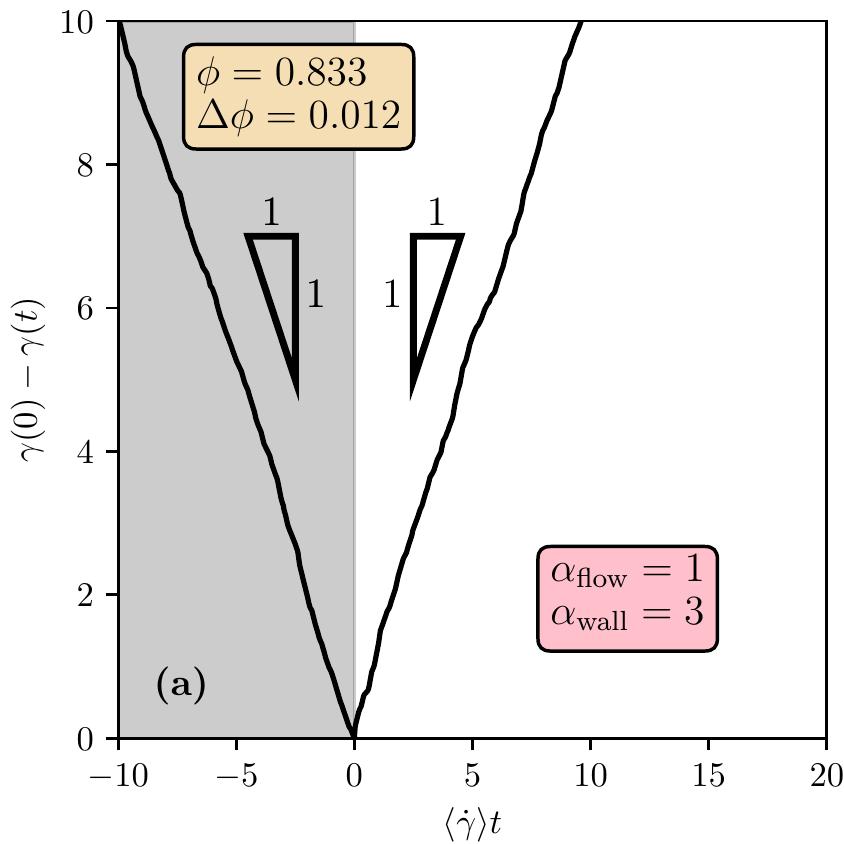}
\includegraphics[width=0.23\textwidth]{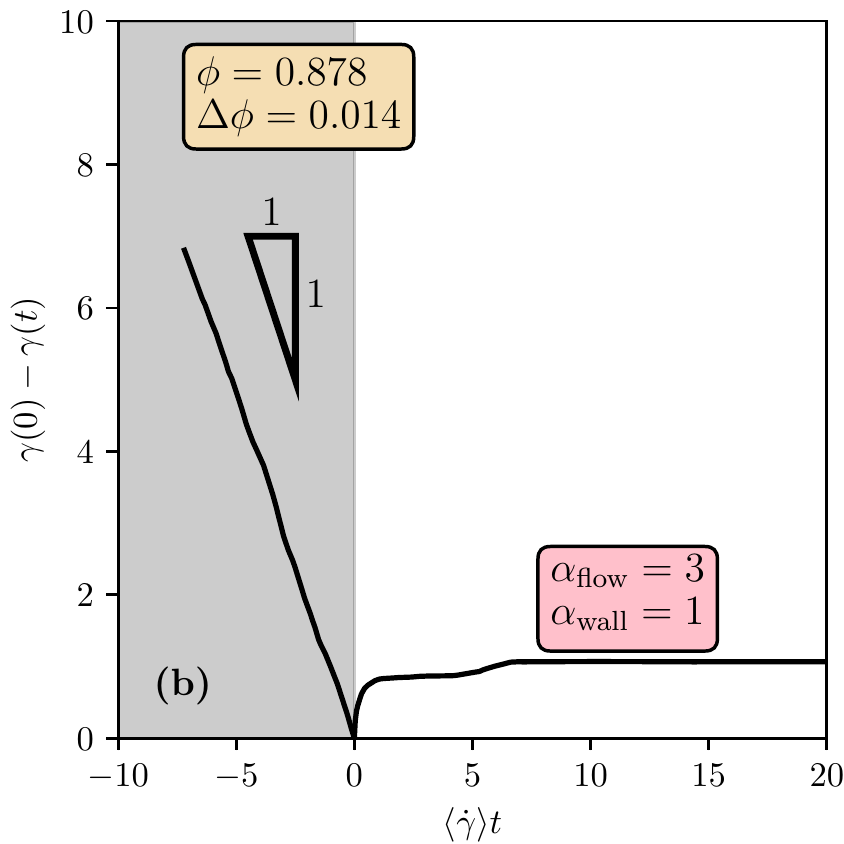}
\caption{ Strain $\gamma$ as a function of the rescaled time $\langle \dot \gamma\rangle t$. {\bf (a)} Ellipse walls with flowing discs and {\bf (b)} disc walls with flowing ellipses.}
 \label{fgr:wall}
\end{figure}
}


\begin{thebibliography}{10}
\bibitem{Andreotti13}
Andreotti B., Forterre Y., and Pouliquen O. \emph{Granular media: between fluid and solid.} (Cambridge University Press, 2013).
\bibitem{Lerner12}
Lerner E., D\"uring G., and Wyart M. A unified framework for non-Brownian suspension flows and soft amorphous solids. PNAS {\bf 109}, 4798-4803 (2012).
\bibitem{Olsson19} 
Olsson P. Dimensionality and viscosity exponent in shear-driven jamming. hys.~Rev.~Lett.~{\bf 122}, 108003 (2019).
\bibitem{Baule18}
Baule A., Morone F., Herrmann H.~J., and Makse H.~A. Edwards statistical mechanics for jammed granular matter. Rev.~Mod.~Phys. {\bf 90}, 015006 (2018).
\bibitem{Denn14}
Denn M.~M.~and Morris J.~M. Rheology of Non-Brownian Suspensions. Annu.~Rev.~Chem.~{\bf 5}, 203-228 (2014).
\bibitem{Guazzelli18}
Guazzelli E. and Pouliquen O. Rheology of dense granular suspensions. J.~Fluid~Mech. {\bf 852}, P1 (2018).
\bibitem{Fielding14}
Fielding S.~M. Shear banding in soft glassy materials. Rep.~Prog.~Phys. {\bf 77}, 102601 (2014). 
\bibitem{Seto13}
Seto R., Mari R., Morris J.~F., Denn M.~M. Discontinuous shear thickening of frictional hard-sphere suspensions. Phys.~Rev.~Lett.~{\bf 111}, 218301 (2013).
\bibitem{Wyart14}
Wyart M. and Cates M. Discontinuous Shear Thickening without Inertia in Dense Non-Brownian Suspensions. Phys.~Rev.~Lett.~{\bf 112}, 098302 (2014).
\bibitem{Dong17}
Dong J. and Trulsson M. Analog of discontinuous shear thickening flows under confining pressure. Phys.~Rev.~Fluids {\bf 2}, 081301(R) (2017).
 \bibitem{Dong20a}
 Dong J. and Trulsson M.  Unifying viscous and inertial regimes of discontinuous shear thickening suspensions. J.~Rheology {\bf 64}, 255-266 (2020).
 \bibitem{Pine05}
Pine D.~J., Gollub J.~P., Brady J.~F., and Leshansky A.~M. Chaos and threshold for irreversibility in sheared suspensions. Nature {\bf 438}, 997-1000 (2005).
 \bibitem{Corte08}
Cort\'e L., Chaikin P.~M., Gollub J.~P., and Pine D.~J. Random organization in periodically driven systems. Nature Physics {\bf 4}, 420-424 (2008).
\bibitem{Taylor00} 
Taylor G.~I. Kinematic Reversibility, in Multi-Media Fluid Mechanics CD-ROM, edited by G. M. Homsy (Cambridge University Press, Cambridge, 2000).
\bibitem{Souzy16}
Souzy M., Pham P., and Metzger B. Taylor's experiment in a periodically sheared particulate suspension. Phys.~Rev.~Fluids {\bf 1}, 042001(R) (2016).
\bibitem{Das20}
Das P.,Vinutha H.~A., and Sastry S. Unified phase diagram of reversible-irreversible, jamming, and yielding transitions in cyclically sheared soft-sphere packings. PNAS {\bf 117}, 10203-10209 (2020).
 \bibitem{Lin15}
Lin N.~Y.~C., Guy B.~M.,Hermes M., Ness C., Sun J., Poon W.~C.~K., and Cohen I. Hydrodynamic and Contact Contributions to Continuous Shear Thickening in Colloidal Suspensions. Phys.~Rev.~Lett.~{\bf 115}, 228304 (2015).
\bibitem{Ness18}
Ness C., Mari R., and Cates M.~E. Shaken and stirred: Random organization reduces viscosity and dissipation in granular suspensions. Science Advances {\bf 3}, eaar3296 (2018).
\bibitem{Dong20b}
Dong J. and Trulsson M. Transition from steady shear to oscillatory shear rheology of dense suspensions. Phys.~Rev.~E~{\bf 102}, 052605 (2020).
\bibitem{Seto19}
Seto R., Singh A., Chakraborty B., Denn M.~M., and Morris J.~F. Shear jamming and fragility in dense suspensions. Granular Matter {\bf 21}, 82 (2019). 
\bibitem{Bi11} 
Bi D., Zhang J., Chakraborty B., and Behringer R.~B. Jamming by shear. Nature {\bf 480}, 355-358 (2011).
\bibitem{Zhao19}
Zhao Y., Bar\'es J., Zheng H.,  Socolar J.~E.~S., and Behringer B. Shear-Jammed, Fragile, and Steady States in Homogeneously Strained Granular Materials. 
Phys.~Rev.~Lett.~{\bf 123}, 158001 (2019).
 \bibitem{SI}
 See online Supplementary Information.
\bibitem{Trulsson18}
Trulsson M. Rheology and shear jamming of frictional ellipses. J.~Fluid~Mech. {\bf 849}, 718-740 (2018).
\bibitem{Berthier09}
Berthier L. and Biroli G. Glasses and Aging, A Statistical Mechanics Perspective on. In: Meyers R. (eds) Encyclopedia of Complexity and Systems Science. Springer, New York, NY (2009).
\bibitem{Nadler18}
Nadler B., Guillard F., and Einav I.. Kinematic Model of Transient Shape-Induced Anisotropy in Dense Granular Flow. Phys.~Rev.~Lett.~{\bf 120}, 198003 (2018).
\bibitem{Zheng14}
Zheng Z., Ni R., Wang F., Dijkstra M., Wang Y., and Han Y.  Structural signatures of dynamic heterogeneities in monolayers of colloidal ellipsoids. Nat. Commun, {\bf 5}, 3829 (2014). 
\bibitem{Nagy17}
Nagy D.~B., Claudin P., B\"orzs\"onyi T., and Somfai E. Rheology of dense granular flows for elongated particles. Phys.~Rev.~E {\bf 96}, 062903 (2017).
 \bibitem{Boyer11}
Boyer F., Guazzelli E., and Pouliquen O. Unifying Suspension and Granular Rheology.  Phys.~Rev.~Lett.~{\bf 107}, 188301 (2011).
\bibitem{Trulsson12}
Trulsson M., Andreotti B., and Claudin P. Transition from the viscous to inertial regime in dense suspensions. Phys.~Rev.~Lett.~{\bf 109}, 118305 (2012).
\bibitem{Amarsid17}
Amarsid, L., Delenne, J.-Y., Mutabaruka, P., Monerie, Y., Perales, F., and Radjai, F. Viscoinertial regime of immersed granular flows. 
Phys.~Rev.~E~{\bf 96}, 012901 (2017).
 \end{thebibliography}

\begin{thebibliography}{10}
\bibitem{Trulsson18}
Trulsson M. Rheology and shear jamming of frictional ellipses. J.~Fluid~Mech. {\bf 849}, 718-740 (2018).
\bibitem{Chwang75}
Chwang, A. and Wu, T. Hydromechanics of low-Reynolds-number flow. Part 2. Singularity method for Stokes flows. J.~Fluid~Mech. {\bf 67}, 787-815 (1975). 
\bibitem{Dong21}
Dong J. and Trulsson M. Oscillatory shear flows of dense suspensions at imposed pressure: Rheology and micro-structure. arXiv:2011.13215. 
 \end{thebibliography}
\end{document}